\documentstyle[pra,aps,twocolumn]{revtex}
\newcommand{\be}{\begin{equation}}
\newcommand{\ee}{\end{equation}}  

\begin{document}
\draft
\title{The Hydrogen Atom in Combined Electric and Magnetic Fields 
with Arbitrary Mutual Orientations}
\author{J\"org Main, Michael Schwacke, and G\"unter Wunner}
\address{Institut f\"ur Theoretische Physik I, Ruhr-Universit\"at Bochum,
D-44780 Bochum, Germany}
\date{August 28, 1997}
\maketitle

\begin{abstract}
For the hydrogen atom in combined magnetic and electric fields we
investigate the dependence of the quantum spectra, classical dynamics,
and statistical distributions of energy levels on the mutual orientation 
of the two external fields.
Resonance energies and oscillator strengths are obtained by exact 
diagonalization of the Hamiltonian in a complete basis set, even far above 
the ionization threshold.
At high excitation energies around the Stark saddle point the eigenenergies
exhibit strong level repulsions when the angle between the fields is varied.
The large avoided crossings occur between states with the same approximately 
conserved principal quantum number, $n$, and this intramanifold mixing of
states cannot be explained, not even qualitatively, by conventional 
perturbation theory.
However, it is well reproduced by an extended perturbation theory which takes 
into account all couplings between the angular momentum and Runge-Lenz vector.
The large avoided crossings are interpreted as a quantum manifestation of
classical {\em intramanifold chaos}.
This interpretation is supported  by both classical {\em Poincar\'e 
surfaces of section}, which reveal a mixed regular-chaotic intramanifold 
dynamics, and the statistical analysis of {\em nearest-neighbor-spacing} 
distributions.

\end{abstract}

\pacs{PACS numbers: 32.60.+i, 03.65.Sq, 05.45.+b, 32.70.Cs}

\section{Introduction}
The effects of external electric and magnetic fields on atomic spectra
is a fundamental question of atomic physics.
However, up to now most investigations have concentrated  on atoms in only 
one of the two fields, or on selected mutual orientations such as parallel
or perpendicular fields.
For example, Rydberg atoms in magnetic fields are nonintegrable
systems and have been shown to be ideally suited for  studies of the quantum
manifestations of classical chaos \cite{Fri89,Has89,Wat93}.
Because of the cylindrical symmetry around the magnetic field axis,
one component of the angular momentum is conserved, and the problem
is nonseparable in two degrees of freedom.
The cylindrical symmetry is broken in combined nonparallel magnetic 
and electric fields, and the system becomes nonseparable in {\em three} 
degrees of freedom.
The special case of perpendicular fields has been investigated both
experimentally \cite{Wie89,Rai91,Rai94} 
and theoretically by quantum \cite{Pau26,Sol83,Mai92} 
and classical \cite{Gou93,Mil94,Mil96,Mil97} methods.
However, the case of arbitrary field orientation is the most general
situation for atoms in uniform external fields, and therefore deserves
appropriate attention.
For the general field arrangement quantum calculations have been performed 
so far for weak external fields in first and second order perturbation 
theory \cite{Pau26,Sol83} and in the regime of very strong magnetic and 
electric fields \cite{Mel93,Fas96}.

In this Paper we investigate the hydrogen atom in external fields with 
arbitrary mutual orientations.
In Sec.\ II we calculate the (numerically) exact quantum mechanical 
eigenenergies and oscillator strengths for a wide energy and field range, 
even far above the ionization threshold (Stark saddle point), by 
diagonalization of the Hamiltonian matrix in a complete set of eigenfunctions.
On the basis of these exact quantum calculations we investigate the 
dependence of resonance energies on the angle between the external fields.
At energies around the ionization threshold we observe large avoided crossings,
which cannot be explained even qualitatively by conventional perturbation 
theory.
These avoided crossings mainly occur between eigenstates of the same
approximately conserved principal quantum number $n$.
The phenomenon is interpreted as a quantum manifestation of 
{\em intramanifold chaos}, which has recently been discovered in classical 
investigations of the hydrogen atom in perpendicular crossed magnetic and 
electric fields \cite{Mil94}.

To verify the interpretation of the avoided crossings as {\em intramanifold 
chaos} we extend, in Sec.\ III, the conventional perturbation theory
\cite{Sol83} by taking into account all couplings between the angular 
momentum and the Runge-Lenz vector.
This extended quantum mechanical perturbation theory can well reproduce,
at least qualitatively, the large avoided crossings between levels of the
same principal quantum number $n$.

Based on the Hamiltonian of the extended perturbation theory we analyze,
in Sec.\ IV, the classical motion of the angular momentum and the Runge-Lenz 
vector by means of Poincar\'e surfaces of section (PSOS).
The PSOS reveal large chaotic fractions of the classical phase space
for those energy-field regions and mutual field orientations where the 
quantum spectra exhibit large avoided crossings.
The classical calculations therefore confirm the interpretation of 
the intra $n$-manifold level repulsion as a quantum manifestation of
intramanifold chaos.

In Sec.\ V we investigate the {\em nearest-neighbor-spacing} distributions 
of states with the same principal quantum number $n$.
The distributions show a Poissonian behavior for parallel fields, where the 
intramanifold dynamics is regular.
For arbitrarily oriented fields the distribution turns to a Brody-type 
distribution, indicating a mixed regular-chaotic intramanifold dynamics, 
in agreement with the classical Poincar\'e surface of section analysis.

\section{Exact quantum calculations}

The electronic motion of the hydrogen atom in uniform magnetic and 
electric fields with arbitrary mutual orientations is nonseparable in three 
degrees of freedom. The Hamiltonian [in atomic units,
$f = F/(5.14 \times 10^9~ {\rm V/cm})$, 
$\gamma = B/(2.35 \times 10^5~ {\rm T})$]
reads
\be
H = \frac{1}{2} {\bf p}^2 - \frac{1}{r} +  \frac{1}{2} \gamma L_z +
     \frac{1}{8} \gamma^2 (x^2 + y^2) + f_\perp x + f_\parallel z \; ,
\label{hamilt}
\ee
where the magnetic field is oriented in the $z$-direction and the electric 
field in the $(x,z)$-plane with $f_\parallel$ and $f_\perp$ the components 
of the electric field parallel and perpendicular to the magnetic field axis, 
respectively. 
We neglect relativistic and spin effects as well as effects due to the finite
nuclear mass, which yield only very small contributions in the energy-field
regions we examine.

For the exact quantum calculations of the hydrogen atom in external fields
with arbitrary mutual orientations we extend the method described in Ref.\ 
\cite{Mai92} for the special case of the hydrogen atom in perpendicular 
magnetic and electric fields $(f_\parallel=0)$.
In perpendicular fields the parity with respect to the $(z=0)$-plane,
$\pi_z$, is an exact quantum number and allows diagonalizing the Hamiltonian
in each subspace $\pi_z=\pm1$ separately.
This symmetry is broken by the parallel component of the electric field
resulting in an increase of the basis size for the numerical diagonalizations
by about a factor of 2.

For the numerical calculations we transform the Hamiltonian
by introducing dilated semiparabolic coordinates
\be
 \mu = \frac{1}{b} \sqrt{r+z} \; , \quad 
 \nu = \frac{1}{b} \sqrt{r-z} \; , \quad
 \phi = \tan^{-1}{\frac{y}{x}} \; ,
\label{semipara}
\ee
where $b$ is a free length scale parameter. The Schr\"odinger equation in 
dilated semiparabolic coordinates then reads
\begin{eqnarray}
& & \Big[ \triangle_\mu +\triangle_\nu -(\mu^2 + \nu^2)
    + b^4 \gamma  (\mu^2 + \nu^2) i{\partial\over\partial\phi} \nonumber\\
& & {} - \frac{1}{4} (b^4 \gamma)^2 \mu^2 \nu^2(\mu^2 + \nu^2)
    - b^6  f_\parallel (\mu^4 - \nu^4) \nonumber\\
& & {} - 2 b^6 f_\perp \mu \nu (\mu^2 + \nu^2) \cos\phi + 4 b^2 \Big] 
    \Psi \nonumber \\
&=& \lambda \left(\mu^2 + \nu^2\right) \Psi \; ,
\label{dila}
\end{eqnarray}
with
\begin{equation}
 \triangle_{\rho} = \frac{\partial^2}{\partial\rho^2}
   + \frac{1}{\rho} \frac{\partial}{\partial\rho} 
   + \frac{1}{\rho^2} \frac{\partial^2}{\partial\phi^2} \; ; \quad
  (\rho =  \mu ~ {\rm or} ~ \nu)  \; ,
\end{equation}
and
\begin{equation}
 \lambda = -(1 + 2 b^4 E) \; .
\end{equation}
The Schr\"odinger equation (\ref{dila}) has the form of two coupled 
two-dimensional harmonic oscillators, and is therefore conveniently 
represented, in matrix form, in terms of the complete set of basis 
functions given by the product of the eigenstates $|N_\rho m\rangle$ 
of the two-dimensional harmonic oscillator, viz.\
\be
| N_\mu N_\nu m \rangle = | N_\mu m \rangle \times | N_\nu m \rangle \; .
\label{basis}
\ee
One is led to a generalized eigenvalue problem with sparse symmetric matrices.
The elements of these matrices can be calculated using the familiar operator 
relations for the two-dimensional harmonic oscillator.
More details of the method are described in \cite{Schw94}.

To account for continuum states we adopt the complex-rotation method 
\cite{Nut69,Rei82}, which is based on the replacement of the coordinate vector
${\bf r}$ with ${\bf r} e^{i \theta}$ in the Hamiltonian and the wave 
function, and has proved very efficient in the calculation of resonances above
the ionization threshold for the hydrogen atom in magnetic fields \cite{Del91}
and in crossed magnetic and electric fields \cite{Mai92}.
By this transformation, hidden resonances of the Hamiltonian in the continuum,
associated with complex eigenvalues, are exposed, while the resonance wave
functions can still be described by the ${\cal L}_2$ integrable basis functions
(\ref{basis}), but with complex arguments.
In our approach, the complex rotation by the angle $\theta$ corresponds to 
the complex dilatation
\be
b = |b| e^{i \frac{\theta}{2}}
\ee
in (\ref{semipara}) and in Schr\"odinger's equation (\ref{dila}).

Numerically a generalized eigenvalue problem with complex symmetric 
non-Hermitian matrices has to be solved, which was achieved by extending 
the {\em spectral transformation Lanczos method} \cite{Eri80} to complex 
matrices.
 From the imaginary part of the complex energies, $E$, we  
obtain the corresponding widths, $\Gamma$, and lifetimes, $T$, 
of the resonances by
\be
 \Gamma = \frac{\hbar}{T} = -2~ {\rm Im} \; E \; .
\ee

As an example we calculated spectra of the hydrogen atom in combined 
magnetic and electric fields with $B=100$ T, $F=50$ kV/cm.
These field strengths are by about an order of magnitude larger than
typical laboratory fields.
However, they allow studying the interesting physical effects at energy
regions with a reasonably low density of states and reduce the computer 
resources (CPU time and storage requirements) necessary for the numerical 
calculation of the spectra.
Bound states and resonances were obtained by numerical
diagonalization of Eq.\ \ref{dila} in a basis set of dimension $\le 6201$.
The complex energies are presented in Fig.\ 1 for an angle 
$\beta=\arctan(f_\parallel/f_\perp)=\pi/4$ between the fields.
The energy of the Stark saddle point, 
$E_{\rm SP}=-2\sqrt{|{\bf f}|}~ {\rm au}=-1368~ {\rm cm}^{-1}$ 
is marked by an arrow in Fig.\ 1.
Below the threshold all eigenvalues are located on the real axis 
(see Fig.\ 1). 
\newpage
\phantom{}
\begin{figure}[htb]
\vspace{6.0cm}
\includegraphics{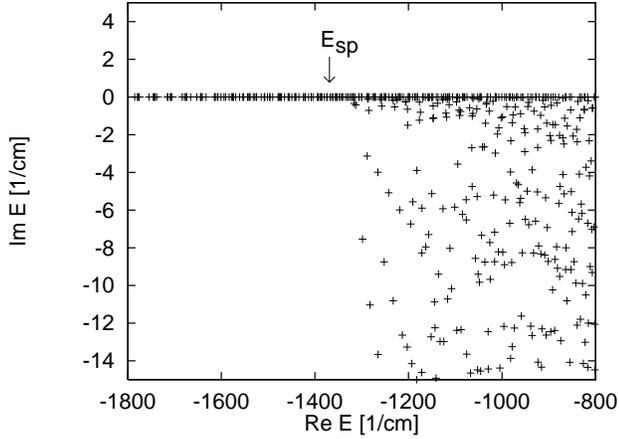}
\caption{\label{fig1} 
Eigenenergies and resonances in the complex energy plane for the hydrogen
atom in magnetic and electric fields ($B=100$ T, $F=50$ kV/cm) of mutual
orientation $\beta=45^\circ$. 
The energy of the Stark saddle point is marked by an arrow.}
\end{figure}
Above threshold, long-lived states close to the real energy axis still exist,
but hidden resonances of the Hamiltonian, associated with complex 
eigenvalues, are exposed by the complex rotation method.

With the energy eigenvalues and eigenvectors of (\ref{dila}) at hand
it is also possible to calculate the cross section for dipole transitions 
from an initial state $\Psi_0$ with  energy $E_0$.
The cross section can be written in the form \cite{Res75}
\begin{figure}[h]
\vspace{8.3cm}
\includegraphics{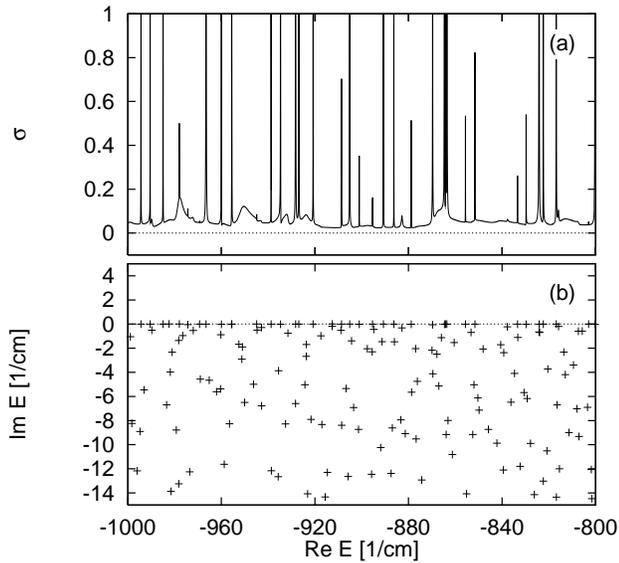}
\caption{\label{fig2} 
(a) Photoabsorption spectrum for the hydrogen atom in magnetic and electric 
fields ($B=100$ T, $F=50$ kV/cm) of mutual orientation $\beta=45^\circ$. 
Transitions $|2p0\rangle \to |\psi_f\rangle$ with light polarized parallel
to the magnetic field axis.
(b) Resonances in the complex energy plane.}
\end{figure}
\begin{equation}
 \sigma(E) = 4 \pi \alpha(E - E_0)~ {\rm Im} \, 
 \left[ \sum_j \frac{\langle\Psi_0|D|\Psi_j(\theta)\rangle^2}
    {E_j(\theta)-E} \right] \; ,
\label{phos}
\end{equation}
where $\Psi_j(\theta)$ are final states at complex energies $E_j(\theta)$,
$D$ denotes the dipole operator for some given polarization, and
$\alpha\approx 1/137$ is the fine-structure constant.
As an example a photoabsorption spectrum for transitions from the initial 
state $|2p0\rangle$ with light polarized parallel to the magnetic field axis
is presented in Fig.\ 2a.
The chosen energy region is the high energy part of Fig.\ 1 and is well 
above the Stark saddle point, $E_{SP}=-1368$ cm$^{-1}$.
The spectrum clearly exhibits sharp peaks of long-lived states superimposed
on broad line shapes of rapidly decaying resonances and a continuous
background.
The complex energy eigenvalues are shown in Fig.\ 2b for comparison.

For atoms in external fields with arbitrary mutual orientations the angle
$\beta$ between the magnetic and electric field axis is an additional free
parameter which allows one to continuously  vary the geometry between the 
extreme situations of parallel and perpendicular fields.
It is therefore interesting to fix the absolute values of the electric and
magnetic field strength and to follow the energy levels as a function of
the angle $\beta$.
A part of such an $(E,\beta)$-diagram is shown in Fig.\ 3 at $B=100$ T, 
$F=50$ kV/cm.
The dotted line marks the energy of the Stark saddle point.
The $(E,\beta)$-diagram exhibits quite complicated patterns which cannot
be explained, not even qualitatively, by conventional perturbation theory
\cite{Sol83}.
%
%\newpage
%\phantom{}
%
\begin{figure}[htb]
\vspace{7.0cm}
\includegraphics{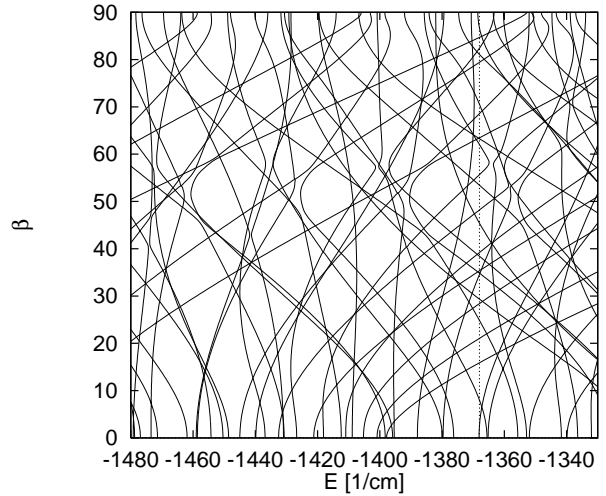}
\caption{\label{fig3} 
Energy-angle diagram for the hydrogen atom in magnetic and electric
fields at $B=100$ T, $F=50$ kV/cm.
$\beta$ is the angle between the two field axis.}
\end{figure}
In particular, many levels undergo large avoided crossings, which are found 
to be most pronounced around angles $40^\circ<\beta<70^\circ$.
The avoided crossings indicate chaos in the underlying classical dynamics.
It is the purpose of the following Sections to investigate and analyze
the patterns observed in Fig.\ 3 in more detail.
This is performed by means of an extended quantum and classical perturbation 
theory, which will allow the interpretation of the avoided crossings 
as effects of {\em intramanifold chaos}.

\section{Extended Quantum Mechanical Perturbation Theory}
The perturbation theory of the hydrogen atom in combined electric and
magnetic fields goes back to the early days of quantum mechanics.
The first order perturbation theory was developed by Epstein \cite{Eps23},
Born \cite{Bor25}, and Pauli \cite{Pau26}.
The second order equations describing the eigenenergies up to the quadratic 
terms in the field strengths were derived by Solov'ev \cite{Sol83}.
The conventional perturbation theory is based classically on the construction 
of adiabatic constants of motion which are related to approximate quantum
numbers in quantum mechanics.
The results of second order perturbation theory agree well with our exact
quantum calculations for weak external fields and at low excitation
energies.
However, at strong perturbations, e.g., at excitation energies around the
Stark saddle point, the conventional perturbation theory
cannot explain the breakdown of approximate quantum numbers and the onset
of chaos in the classical dynamics.
Therefore the large avoided crossings which have been observed, e.g., 
in Fig.\ 3 are not reproduced by the conventional perturbation theory.
In the following we will briefly review the conventional perturbation theory
and then generalize the method by taking into account {\em all} couplings 
between states within the same $n$-manifold up to second order in the external 
fields.
The concept does not assume the existence of a complete set of approximate 
quantum numbers and therefore can describe and explain the phenomenon of 
intramanifold chaos.

The perturbation series for the hydrogen atom in external magnetic and 
electric fields up to second order in the field strengths reads
\begin{equation}
 H_n = - \frac{1}{2 n^2} + V_1 + V_2 + W
\label{Hn}
\end{equation}
with
\begin{equation}
 V_1 = {1\over 2}\gamma L_z + {\bf f}\cdot {\bf r}
\label{V1}
\end{equation}
the paramagnetic and linear Stark terms,
\begin{equation}
 V_2 = {1\over 8}\gamma^2\rho^2
\label{V2}
\end{equation}
the diamagnetic term, and
\begin{equation}
 W = ({\bf f} \cdot {\bf r}) G_n ({\bf f} \cdot {\bf r})
\label{W}
\end{equation}
the quadratic Stark effect, with $G_n$ the Green's function of the field free
hydrogen atom, $H_0$.
The operators $V_1$, $V_2$, and $W$ can now be replaced within each 
$n$-manifold by operator identities in terms of the angular momentum
\begin{equation}
 {\bf L} = {\bf r} \times {\bf p}
\label{L_def}
\end{equation}
and the Runge-Lenz vector
\begin{equation}
 {\bf A} = {1\over\sqrt{-2H_0}} \left[{1\over2} \left({\bf p} \times {\bf L}
     - {\bf L} \times {\bf p}\right) - {{\bf r}\over r} \right]
\label{A_def}
\end{equation}
or in terms of the linear combinations of these vectors
\begin{equation}
 {\bf I}_{1,2} = {1\over2}({\bf L} \pm {\bf A}) \; .
\label{I_def}
\end{equation}
The vectors ${\bf I}_1$ and ${\bf I}_2$ commute with each other and their 
components obey the commutation relations for angular momentum operators,
\begin{eqnarray}
 \left[I_{1j},I_{2k}\right] &=& 0 \; , \nonumber \\
 \left[I_{1j},I_{1k}\right] &=& i \epsilon_{jkl} I_{1l} \; , \nonumber \\
 \left[I_{2j},I_{2k}\right] &=& i \epsilon_{jkl} I_{2l} \; .
\label{I_comm_qm}
\end{eqnarray}
Both vectors have the same norm, which depends on the principal quantum number,
\begin{equation}
 |{\bf I}_1|=|{\bf I}_2|= {1\over 2}\sqrt{n^2-1} \; .
\label{I_norm_qm}
\end{equation}
The operator identity for $V_1$ in a given $n$-manifold then reads \cite{Pau26}
\be
 V_1 = {\vec \omega}_1 \cdot {\bf I}_1 + {\vec \omega}_2 \cdot {\bf I}_2
\label{V1_id}
\ee
with
\be
 {\vec \omega_{1,2}} = \frac{1}{2} \left({\vec\gamma} \mp 3 n {\vec f}\right)
\label{omega_def}
\ee
as is illustrated in Fig.\ 4.
Because ${\bf I}_1$ and ${\bf I}_2$  independently fulfill the commutation 
algebra of two angular momenta, the operator $V_1$ in (\ref{V1_id}) can be 
quantized immediately to obtain the energy correction in first order 
perturbation theory
\begin{eqnarray}
 E^{(1)}_{nn'n''} = \omega_1 n' + \omega_2 n''
\label{E1}
\end{eqnarray}
with $\omega_{1,2}=|{\vec \omega}_{1,2}|$ and quantum numbers 
$n',n'' = -(n-1)/2, -(n-3)/2, \dots, +(n-1)/2$.
Classically, ${\bf I}_1$ and ${\bf I}_2$ turn around the vectors
${\vec \omega}_1$ and ${\vec \omega}_2$, whose absolute values 
are the frequencies of the secular motion (see Fig.\ 4).

The operator identities for $V_2$ and $W$ are the starting point for
the second order perturbation theory and have been derived by Solov'ev 
\cite{Sol83},
\begin{eqnarray}
\label{V2_id}
 V_2 &=& \frac{n^2 \gamma^2}{16} (n^2 + 3 + L_z^2 + 4 {\bf A}^2 - 5 A_z^2)
  \; , \\
\label{W_id}
 W &=& - \frac{n^4 f^2}{16} (5n^2 + 31 + 24 {\bf L}^2 -21 L_f^2 + 9 A_f^2) \; ,
\end{eqnarray}
with $L_f$ and $A_f$ the projections of the angular momentum and Runge-Lenz
vector on the electric field axis.

Except for the special case of perpendicular fields where we have
$\omega_1 = \omega_2 = \frac{1}{2}\sqrt{\gamma^2+9n^2f^2}$, the degeneracy 
of the $n^2$ states belonging to each $n$-manifold
is already completely destroyed in first order perturbation 
\newpage
\phantom{}
\begin{figure}[hbt]
\vspace{6.0cm}
\includegraphics{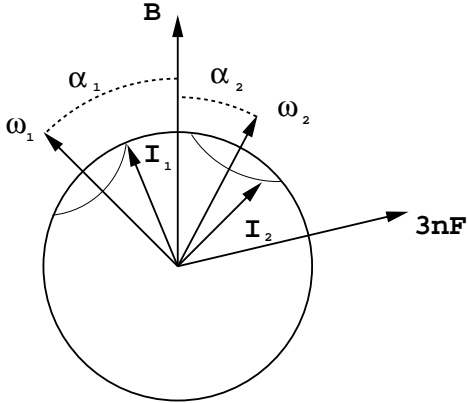}
\caption{\label{fig4} 
Schematic view of the vectors $\vec\omega_1$, $\vec\omega_2$ and 
${\bf I}_1$, ${\bf I}_2$ which are used in perturbation theory.}
\end{figure}
\noindent
theory.
In conventional perturbation theory, the second order energy correction,
$E_{nn'n''}^{(2)}$ is given as the expectation value of $V_2+W$ in the 
eigenstate $|nn'n''\rangle$,
\begin{eqnarray}
 && E_{nn'n''}^{(2)} = \langle nn'n''|V_2+W|nn'n''\rangle \nonumber \\
 &=& -\frac{n^4 f^2}{16}[ 17 n^2 + 19 - 12\, (n'^2+n'n''\cos(\alpha_1
               +\alpha_2) \nonumber \\
 &&  {}+n''^2)] + \frac{n^2 \gamma^2}{48}[7 n^2 + 5 + 4 n'n''\sin\alpha_1
     \sin\alpha_2  \nonumber \\
 &&  {}+(n^2 - 1)(\cos^2 \alpha_1 + \cos^2 \alpha_2)
            - 12\, (n'^2 \cos^2 \alpha_1 \nonumber \\
 &&  {}-n'n'' \cos\alpha_1 \cos\alpha_2 + n''^2 \cos^2 \alpha_2)] \; ,
\label{E2_conv}
\end{eqnarray}
where $\alpha_1$ and $\alpha_2$ are the angles between the magnetic field 
axis and the vectors $ {\vec \omega}_1 $ and $ {\vec \omega}_2 $, respectively
(see Fig.\ 4).
The energy eigenvalues are finally obtained as
$E_{nn'n''}=-1/2n^2+E_{nn'n''}^{(1)}+E_{nn'n''}^{(2)}$.
Note that Eq.\ \ref{E2_conv} is not valid for perpendicular fields because
in this case the degeneracy of the $n$-manifolds is not completely destroyed 
in first order perturbation theory \cite{Sol83,Bra84}.
However, the qualitative results of the following discussion are not changed
when the conventional perturbation theory for nearly perpendicular fields 
is applied instead of Eq.\ \ref{E2_conv} \cite{Schw94}.

The results of the conventional second order perturbation theory for
non-perpendicular fields are compared to the exact quantum calculations 
in Fig.\ 5.
The dashed lines in Figs.\ 5a and 5b are the levels of manifolds $n=9$ 
and $n=10$, respectively.
To simplify the identification the exact levels with the corresponding
principal quantum numbers are highlighted by solid lines while other
levels are drawn by dotted lines.
Although the identification is possible, the comparison reveals remarkable
discrepancies between the results of conventional perturbation theory and the 
exact quantum results in this energy regime.
In the conventional perturbation theory all level crossings are exact
and cannot reproduce the large avoided crossings discussed above.
Another interesting
\newpage
\phantom{}
\begin{figure}[htb]
\vspace{9.2cm}
\includegraphics{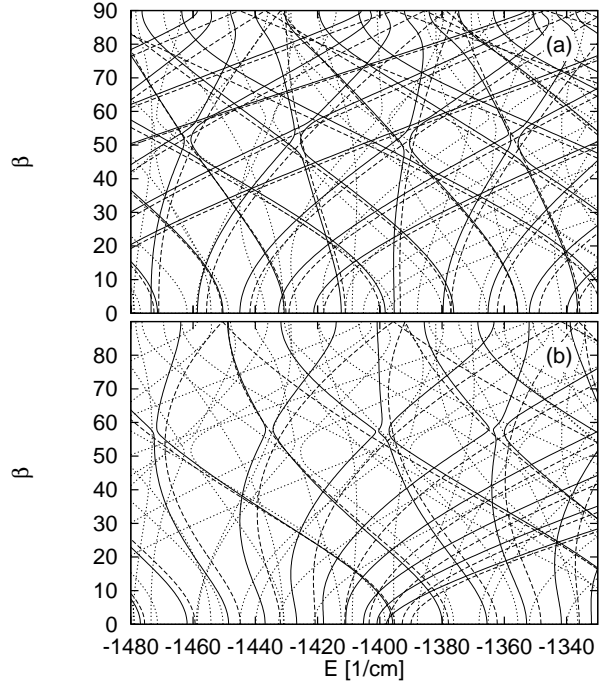}
\caption{\label{fig5}
Exact energy eigenvalues (solid lines) compared with energy values calculated 
by conventional second order perturbation theory (dashed lines). 
(a) Principal quantum number $n=9$; (b) $n=10$.}
\end{figure}
\noindent
result of the comparisons in Fig.\ 5 is that
the large avoided crossings do not occur between levels with different
principal quantum numbers.
Therefore they cannot be interpreted as an $n$-mixing effect, which has been 
identified as the origin of level repulsion in the hydrogen atom in magnetic 
or parallel magnetic and electric fields.
The level repulsions in Fig.\ 5 solely occur between levels of the same 
$n$-manifold and therefore must be interpreted as a mixing between states
$|nn'n''\rangle$ with different quantum numbers $n'$ and $n''$ but with
the same principal quantum number $n$.
The breakdown of the quantum numbers $n'$ and $n''$ is classically related
to {\em intramanifold chaos}, which will be discussed in Sec.\ IV.

We now demonstrate that the avoided crossings are an effect
of intramanifold mixing of states.
The conventional perturbation theory is based on the assumption that the 
external fields are weak, i.e., the linear terms in the external fields, 
$V_1$, and the quadratic terms, $V_2+W$, can be treated separately as first 
and second order perturbations.
We now extend the conventional perturbation theory by diagonalizing the 
complete perturbation operator $V_1 + V_2 + W$ in Eq.\ \ref{Hn} within
a given $n$-manifold.
As a basis set we choose the parabolic states, i.e., simultaneous eigenstates
of the Coulomb Hamiltonian, $H_0$, and the $z$ components of the angular 
momentum and Runge-Lenz vectors, $L_z$ and $A_z$.
By substituting the angular momentum and Runge-Lenz vector in Eqs.\ 
\ref{V1_id}, \ref{V2_id}, and \ref{W_id} with ${\bf L}={\bf I}_1+{\bf I}_2$
and ${\bf A}={\bf I}_1-{\bf I}_2$ and applying the commutator algebra
for the operators ${\bf I}_1$ and ${\bf I}_2$ it is a straightforward task
to compute the required matrix elements of $V_1 + V_2 + W$.
The energy eigenvalues are obtained by numerical diagonalization of the
perturbation matrix of dimension $n^2 \times n^2$.

The results of the extended perturbation theory are presented in Fig.\ 6
for parts of the manifolds $n=9$ and $n=10$.
The exact quantum levels (highlighted solid and dotted lines) are the same 
as in Fig.\ 5.
\begin{figure}[htb]
\vspace{9.3cm}
\includegraphics{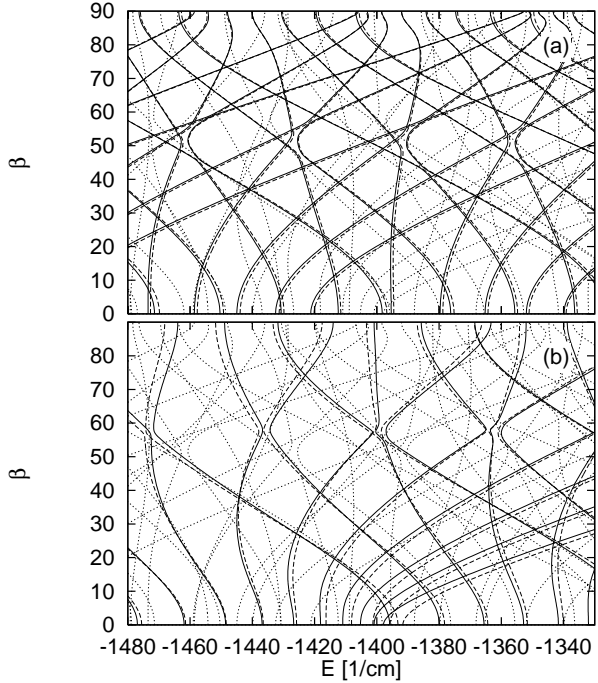}
\caption{\label{fig6} 
Same as Fig.\ 5 but for the comparison between the exact eigenenergies
(solid lines) and the results of extended perturbation theory (dashed lines).
(a) $n=9$; (b) $n=10$.}
\end{figure}
The dashed lines are the eigenvalues from the diagonalization of the complete 
perturbation matrix (\ref{Hn}).
They do not totally agree with the exact levels, but this cannot be expected
at the high excitation energies close to the Stark saddle point.
However, Fig.\ 6 qualitatively exhibits the same level dynamics for both
the solid and dashed lines for all mutual field orientations, i.e., 
from parallel fields $(\beta=0)$ to perpendicular fields $(\beta=\pi/2)$.
In particular, the avoided crossings between levels of the same $n$-manifold
are well reproduced within the extended perturbation theory, and this
strongly supports the interpretation of the observed level repulsion as
an effect of intramanifold mixing.
The $n$-mixing of states will become important only at higher energies far
above the ionization threshold.

\section{Classical Intramanifold Dynamics}
The hydrogen atom in magnetic and electric fields with arbitrary mutual 
orientations is nonintegrable in {\em three} degrees of freedom.
However, for weak external fields and low excitation energies three
approximate quantum numbers $n$, $n'$, and $n''$ exist.
In classical mechanics these quantum numbers are related to three constants 
of the motion, viz.\ the length of the vectors ${\bf I}_1$ and ${\bf I}_2$,
\begin{equation}
 |{\bf I}_1|=|{\bf I}_2|={1 \over 2} n
\label{I_norm_cl}
\end{equation}
(note the slight difference to the quantum expression (\ref{I_norm_qm}))
and their projections on the axis ${\vec \omega}_1$ and 
${\vec \omega}_2$ defined in Eq.\ \ref{omega_def}.
${\bf I}_1$ and ${\bf I}_2$ turn around the vectors ${\vec \omega}_1$ 
and ${\vec \omega}_2$ with constant frequencies (see Fig.\ 4), i.e., 
the classical dynamics is completely regular.
In the extended perturbation theory discussed in the previous Section, 
states $|nn'n''\rangle$ with the same principle quantum number $n$ but
with different quantum numbers $n'$ and $n''$ are mixed, i.e., {\em two} 
approximate quantum numbers are destroyed.
The breakdown of two constants of motion can lead to a chaotic
secular motion of the vectors ${\bf I}_1$ and ${\bf I}_2$ even when the
principal quantum number, $n$, is still conserved.
This phenomenon is called {\em intramanifold chaos} and has been discovered
in the hydrogen atom in perpendicular crossed magnetic and electric fields
\cite{Mil94}.
Here we study the classical intramanifold dynamics of the hydrogen atom
in external fields with arbitrary mutual orientations, and are especially
interested in the dependence of the classical dynamics on the angle between
the external fields.
The quantum calculations in Sec.\ III  revealed large avoided crossings
at angles $40^\circ<\beta<70^\circ$.
If the level repulsion is the quantum counterpart of classical intramanifold
chaos, the classical dynamics should be particularly irregular at those field 
orientations.

To investigate the intramanifold dynamics we solve the equations of motion
for the secular motion of the two vectors ${\bf I}_1$ and ${\bf I}_2$ defined
in (\ref{I_def}).
By expressing the angular momentum and Runge-Lenz vector in the Hamiltonian
(\ref{Hn}) in terms of ${\bf I}_1$ and ${\bf I}_2$, scaled with the principal
quantum number, viz.\ ${\bf I}_{1,2} \to {\bf I}_{1,2}/n$, we obtain the 
classical Hamiltonian valid within a given $n$-manifold,
\begin{eqnarray}
 & & {\cal H} \equiv 2n^2 H_n = 2n^2E \nonumber \\
 &=& -1 + n^3\gamma(I_{1z}+I_{2z}) \nonumber \\
 &-& 3n^4f[\sin\beta(I_{1x}-I_{2x}) + \cos\beta (I_{1z}-I_{2z})] \nonumber \\
 &+& {1\over8}(n^3\gamma)^2[3-4(I_{1z}^2+I_{2z}^2-I_{1z}I_{2z})
          - 8I_{1x}I_{2x} - 8I_{1y}I_{2y}] \nonumber \\
 &-& {1\over8}(n^4f)^2\{17-12\sin^2\beta(I_{1x}^2+I_{2x}^2)
          - 12 \cos^2\beta(I_{1z}^2+I_{2z}^2) \nonumber \\
 &+& (48-60\sin^2\beta)I_{1x}I_{2x} + 48I_{1y}I_{2y}
          + (48-60\cos^2\beta)I_{1z}I_{2z} \nonumber \\
 &-& 12\sin\beta\cos\beta[5(I_{1x}I_{2z}+I_{1z}I_{2x})
          + 2(I_{1x}I_{1z}+I_{2x}I_{2x})]\} \; . \nonumber \\ {}
\label{H_class}
\end{eqnarray}
The classical equations of motions are now obtained from the Hamiltonian
(\ref{H_class}) by the Poisson brackets
\begin{eqnarray}
 \dot {\bf I}_1 = [{\bf I}_1,{\cal H}] \; , \nonumber \\
 \dot {\bf I}_2 = [{\bf I}_2,{\cal H}] \; , 
\end{eqnarray}
which can be solved with the help of the elementary Poisson brackets for
the components of ${\bf I}_1$ and ${\bf I}_2$,
\begin{eqnarray}
 \left[I_{1j},I_{2k}\right] &=& 0 \; , \nonumber \\
 \left[I_{1j},I_{1k}\right] &=& \epsilon_{jkl} I_{1l} \; , \nonumber \\
 \left[I_{2j},I_{2k}\right] &=& \epsilon_{jkl} I_{2l} \; .
\label{I_comm_cl}
\end{eqnarray}
The classical intramanifold dynamics depends on the energy and field strengths
scaled with the principal quantum number, $n^2E$, $n^3\gamma$, and $n^4f$,
and on the angle, $\beta$, between the fields.
The classical trajectories are computed in the six-dimensional space 
$\{{\bf I}_1(t), {\bf I}_2(t)\}$.
However, the norm of ${\bf I}_1$ and ${\bf I}_2$ is conserved 
(\ref{I_norm_cl}), and therefore the effective phase space is four-dimensional.
A convenient parameterization of the four-dimensional phase space are
the projections of ${\bf I}_1$ and ${\bf I}_2$ on the axis 
${\vec \omega}_1$ and ${\vec \omega}_2$ (see Fig.\ 4), called 
$J_{1z}$ and $J_{2z}$ in what follows, and the polar angles $\phi_1$ and
$\phi_2$ of ${\bf I}_1$ and ${\bf I}_2$ with respect to the 
${\vec \omega}_{1,2}$ axis.
Note that $\{J_{1z},\phi_1\}$ and $\{J_{2z},\phi_2\}$ are action-angle 
variables and $J_{1z}$ and $J_{2z}$ are constants of motion in the limit 
of weak non-perpendicular external fields.

A common way of visualizing  the classical dynamics of a system is the 
method of Poincar\'e surfaces of section (PSOS).
In a system with $n$ degrees of freedom the PSOS is in general a 
$(2n-2)$-dimensional subspace of the $(2n)$-dimensional phase space.
Elliptic fixpoints surrounded by torus structures indicate regular classical
dynamics while chaotic motion is indicated by stochastic layers in the PSOS.
For the analysis of the intramanifold dynamics we use the action-angle
variable of the weak field limit and define the intersections with the
PSOS by $\phi_2=0$.
The two-dimensional PSOS is now defined by the doublet $\{J_{1z},\phi_1\}$, 
while $J_{2z}$ is given implicitly via the conservation of energy.
\begin{figure}[htb]
\vspace{7.5cm}
\includegraphics{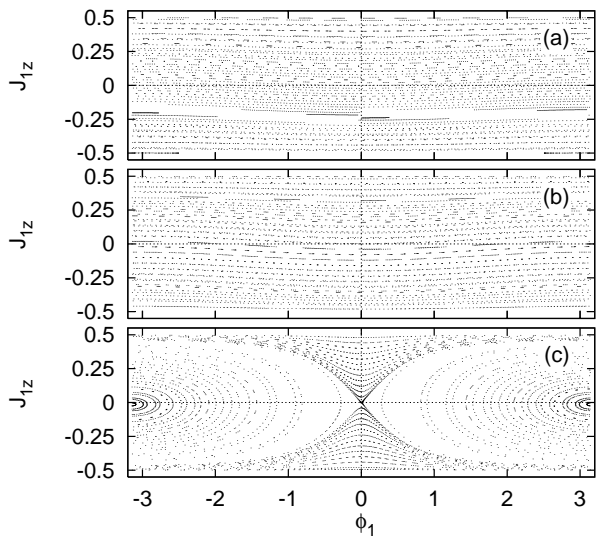}
\caption{\label{fig7} 
Poincar\'e surfaces of section for the classical intramanifold dynamics 
with parameters $n^3 \gamma=0.053$, $n^4 f=0.0061$, $n^2E=-0.5$.
Mutual field orientation (a) $\beta = 0^\circ$, (b) $\beta = 45^\circ$, 
(c) $\beta = 90^\circ$.}
\end{figure}
As an example we calculated PSOS for very weak external fields, 
$n^3\gamma=0.053$, $n^4f=0.0061$, which is related to the manifold $n=5$ 
at the field strengths $B=100$ T, $F=50$ kV/cm chosen for the quantum 
calculations in Secs.\ II and III.
Figs.\ 7a and 7b present the results for mutual field orientations $\beta=0$
(parallel fields) and $\beta=\pi/4$, respectively.
Qualitatively both PSOS look the same, i.e., they exhibit nearly parallel
lines with $J_{1z}\approx{\rm const}$.
This is a verification of the fact that $J_{1z}$ is a constant of motion in
the limit of weak external non-parallel fields.
For the special case of perpendicular fields ($\beta=\pi/2$) the PSOS have
a qualitatively different appearance (Fig.\ 7c).
This is a consequence of the fact that in perpendicular fields the 
degeneracy is not completely destroyed in first order perturbation theory
($\omega_1=\omega_2$ in Eq.\ \ref{E1}), and therefore $J_{1z}$ is no constant
of motion.
However, Fig.\ 7 clearly reveals that the classical intramanifold dynamics is
completely regular in the weak field limit for all mutual field orientations.

The situation changes when we look at the PSOS for stronger interaction
of the hydrogen atom with the external fields.
To search for classical intramanifold chaos we analyzed the dynamics at
parameters $n^3\gamma=0.74$, $n^4f=0.2$, $n^2E=-0.5$, which is related to 
the manifold 
\begin{figure}[htb]
\vspace{11.8cm}
\includegraphics{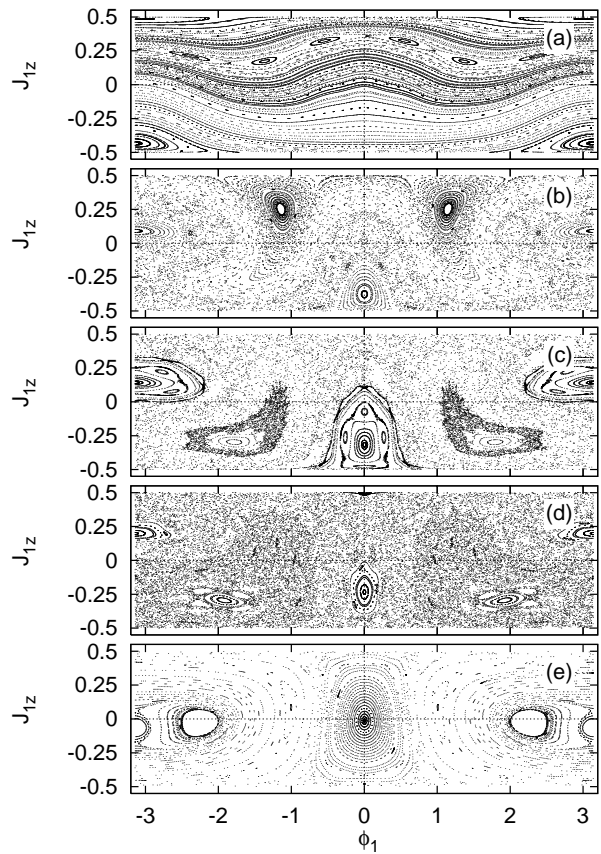}
\caption{\label{fig8}
Same as Fig.\ 7 but with parameters $n^3\gamma=0.74$, $n^4f=0.2$, 
$n^2E=-0.5$ and mutual field orientation (a) $\beta = 20^\circ$, 
(b) $\beta = 58^\circ$, (c) $\beta = 60^\circ$, (d) $\beta = 62^\circ$, 
(e) $\beta = 90^\circ$.}
\end{figure}
\noindent
$n=12$ at the field strengths $B=100$ T, $F=50$ kV/cm.
The Poincar\'e surfaces of section are presented in Fig.\ 8 for various
mutual orientations of the external fields.
In the region $0\le\beta<40^\circ$ the torus structures 
$J_{1z}\approx{\rm const}$ observed at weak fields (Figs.\ 7a and 7b)
are strongly distorted (see Fig.\ 8a at $\beta=20^\circ$).
However, the tori are not destroyed, i.e., the classical intramanifold 
dynamics is mostly regular apart from tiny stochastic regions around
hyperbolic fix points.
Note that no intramanifold level repulsion was observed for those
field orientations in the quantum calculations of Secs.\ II and III.

The onset of strong classical intramanifold chaos is discovered at
mutual field orientations $40^\circ<\beta<70^\circ$.
Examples at $\beta=58^\circ$, $\beta=60^\circ$, and $\beta=62^\circ$, are
given in Figs.\ 8b, 8c, and 8d, respectively.
They clearly exhibit mixed regular-chaotic classical dynamics with large
or even dominating stochastic, i.e., chaotic phase space regions.
Fig.\ 8 also shows a sensitive dependence of the classical dynamics on
the angle between the external fields.
Even small deviations $\Delta\beta=2^\circ$ of the mutual field orientation
result in significant changes of the PSOS.
For example at $\beta = 58^\circ$ (Fig.\ 8b) there exists a regular region 
at the top of the PSOS, which vanishes at $\beta = 60^\circ$ in Fig.\ 8c. 
At that angle the PSOS exhibits two large irregular areas which are
dynamically separated. 
The dynamical barrier is resolved in Fig.\ 8d where only a few stability 
islands are embedded in one large irregular region.
The field arrangement where strong classical intramanifold chaos is observed
($40^\circ<\beta<70^\circ$) coincides with the region where large avoided
crossings between quantum levels with the same principal quantum number, $n$,
were found in Fig.\ 6.
This strongly supports the interpretation of the level repulsion as a
quantum manifestation of intramanifold chaos.

If the angle between the fields is further increased ($\beta>70^\circ$)
the intramanifold dynamics becomes more and more regular again.
The PSOS for the special case of perpendicular fields ($\beta=90^\circ$) 
is presented in Fig.\ 8e.
Chaotic regions exist in agreement with previous classical investigations 
of the crossed field atom \cite{Mil94}, however, the PSOS is clearly 
dominated by regular torus structures.

\section{Intramanifold Level Statistics}
It is well established that the nearest-neighbor-spacing distribution (NNS)
of integrable quantum systems is given, after unfolding the spectra to unit
mean level spacing $\langle s\rangle=1$, by a Poisson distribution
\begin{equation}
 P_{\rm Poisson}(s) = \exp\left[-s\right] \; ,
\label{P_Poisson}
\end{equation}
while quantum systems with a fully chaotic (ergodic) underlying classical
dynamics are characterized by the Wigner distribution
\begin{equation}
 P_{\rm Wigner}(s) = {\pi s \over 2} \exp\left[-{\pi\over4}s^2\right]
\label{P_Wigner}
\end{equation}
obtained from random matrix theory \cite{Boh84,Haa91}.
In systems with a mixed regular-chaotic classical dynamics the  
nearest-neighbor-spacing distribution can be phenomenologically described by 
the Brody distribution \cite{Bro81}
\begin{equation}
 P_{\rm Brody}(s;q) = (q+1) \alpha s^q \exp\left[-\alpha s^{q+1}\right]
\label{P_Brody}
\end{equation}
where
\begin{equation}
 \alpha=\left[\Gamma\left({q+2}\over {q+1}\right)\right]^{q+1}
\end{equation}
and $q$ is a parameter which interpolates between the Poisson distribution 
($q=0$) and the Wigner distribution ($q=1$) and is roughly related to the 
percentage of chaotic phase space volume of the underlying classical system.

Here we do not intend to compare nearest-neighbor-spacing distributions
with the complete three-dimensional classical dynamics of the hydrogen atom
in external magnetic and electric fields with arbitrary mutual orientations.
Instead, we want to search for fingerprints of {\em intramanifold quantum 
chaos} in the level statistics.
We therefore do not investigate the NNS distribution of all levels of a
given spectrum but analyze separately the spacings of neighboring states
with the same principal quantum number, $n$.
For direct comparisons with the PSOS of the classical intramanifold dynamics
(see Sec.\ IV) the spacing distributions should be obtained at constant 
parameters $n^3\gamma$, $n^4f$, and $n^2E$ for the scaled field strengths 
and energy.
In order to get a reliable statistics for the problem we have to calculate 
a sufficiently large set of eigenvalues.
To fulfill these conditions we calculated eigenenergies at constant scaled 
field strengths $n^3\gamma$ and $n^4f$ by means of the extended second order
perturbation theory derived in Sec.\ III.
The Hamiltonian (\ref{Hn}) was diagonalized numerically for all $n$-manifolds
from $n=50$ to $n=60$.
The resulting spectra were unfolded separately to mean level spacing 
$\langle s\rangle=1$ \cite{Boh84}.
For the NNS distributions we included only eigenstates whose scaled energies 
deviate by at most $0.1n^3\gamma$ from the given value of $n^2E$.
This ensures that the scaled energy is approximately constant and the
NNS distributions can be compared to the classical dynamics, viz.\ the PSOS,
at a given constant energy.

As an example, Fig.\ 9 presents the NNS distribution at field parameters 
$n^3\gamma=0.31$, $n^4f=0.064$, $\beta=45^\circ$, and for approximately
constant scaled energy $n^2E \approx -0.5$.
These parameters represent the situation of the manifold $n=9$ at $B=100$ T
and $F=50$ kV/cm in Fig.\ 6a.
However, to increase the density of states for a more reliable statistics
we analyzed the manifolds $n=50$ to $n=60$ at correspondingly rescaled
energies and field strengths.
The histogram in Fig.\ 9 is close to a Poisson distribution (\ref{P_Poisson})
indicating a regular intramanifold dynamics.
An even better fit is obtained by a Brody distribution (\ref{P_Brody}) with 
$q=0.08$ (dashed line in Fig.\ 9).
The deviations from the Poisson distribution can be interpreted as the
onset of intramanifold quantum chaos.
The decrease of 
\newpage
\phantom{}
\begin{figure}[htb]
\vspace{4.2cm}
\includegraphics{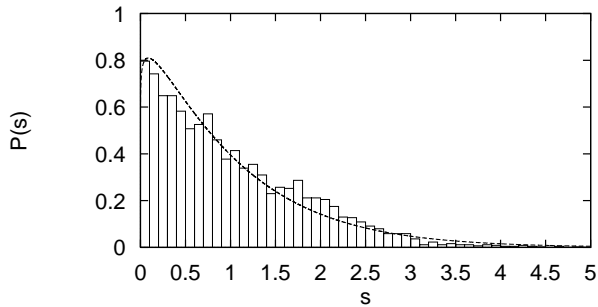}
\caption{\label{fig9} 
Intramanifold nearest-neighbor-spacing distribution for the parameters
$n^3\gamma=0.31$, $n^4f=0.064$, $n^2E\approx -0.5$, $\beta=45^\circ$.
The dashed line is a Brody distribution with $q=0.08$.}
\end{figure}
\noindent
the probability $P(s)$ at small spacings $s$ is related
to the observation of level repulsions in Fig.\ 6a.
For a more pronounced verification of intramanifold quantum chaos in
NNS distributions we investigated quantum spectra at parameters
$n^3\gamma=0.74$, $n^4f=0.2$, $n^2E \approx -0.5$, i.e., the same 
parameters as for the analysis of the classical intramanifold dynamics
in Fig.\ 8 of Sec.\ IV.
The NNS distributions for the same set of five different mutual field 
orientations between $\beta=20^\circ$ and $\beta=90^\circ$ are presented
in Fig.\ 10.
The histograms indicate a predominately regular intramanifold dynamics 
at $\beta=20^\circ$ (Fig.\ 10a) and $\beta=90^\circ$ (Fig.\ 10e) with
Brody parameters $q=0.1$ and $q=0.05$, respectively, and this is in
excellent agreement with the regular torus structures in the PSOS of
Figs.\ 8a and 8e.
The NNS distribution changes dramatically at mutual field orientations
around $\beta=60^\circ$ (see Figs.\ 10b -- 10d).
The histograms turn into Brody distributions with much higher Brody 
parameters, up to $q=0.45$, indicating a large irregular part of the 
underlying classical intramanifold dynamics.
Again the behavior of the NNS distributions is in very good agreement with 
the PSOS of Figs.\ 8b -- 8d, which exhibit large stochastic regions at field 
orientations around $\beta=60^\circ$.
Another interesting aspect which can be observed in Fig.\ 10 is the sensitive 
dependence of the NNS distributions on small variations of the mutual
orientation between the external fields around $\beta\approx 60^\circ$.
The Brody parameter increases from  $q=0.3$ at $\beta=58^\circ$ 
(Fig.\ 10b) to $q=0.45$ at $\beta=60^\circ$ (Fig.\ 10c) and then decreases 
to $q=0.2$ at $\beta=62^\circ$ (Fig.\ 10d).
The rapid changes of the Brody parameter reflect the sensitive dependence
of the classical intramanifold dynamics on the field orientation discovered
in the PSOS (see Figs.\ 8b -- 8d).

\section{Conclusion}
We have investigated the hydrogen atom in electric and magnetic fields
with arbitrary mutual orientations by means of exact quantum calculations,
an extended 
\newpage
\phantom{}
\begin{figure}[htb]
\vspace{11.5cm}
\includegraphics{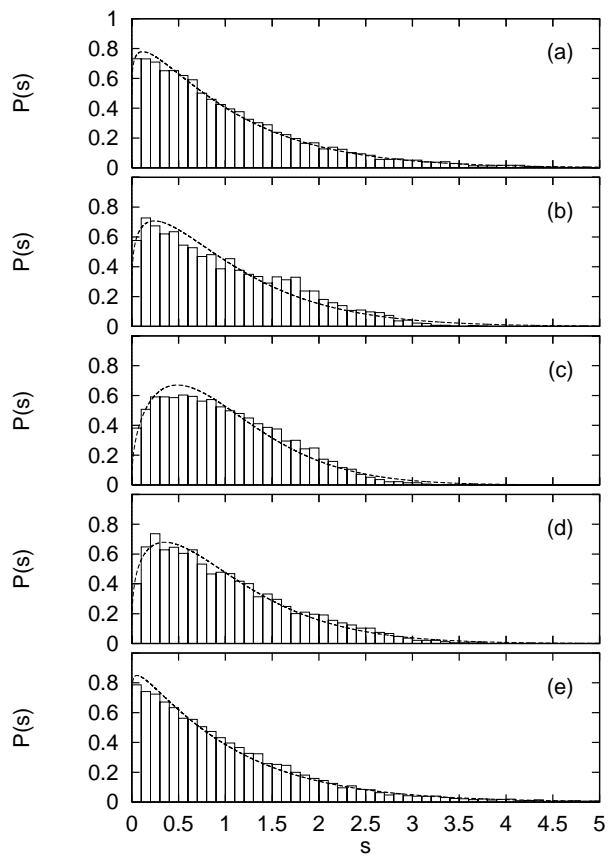}
\caption{\label{fig10} 
Intramanifold nearest-neighbor-spacing distributions for the parameters
$n^3\gamma=0.74$, $n^4f=0.2$, $n^2E \approx -0.5$ and mutual field 
orientations (a) $\beta = 20^\circ$, (b) $\beta = 58^\circ$, 
(c) $\beta = 60^\circ$, (d) $\beta = 62^\circ$, (e) $\beta = 90^\circ$.
The dashed lines are fitted Brody distributions with 
(a) $q=0.1$, (b) $q=0.2$, (c) $q=0.45$, (d) $q=0.3$, (e) $q=0.05$.}
\end{figure}
\noindent
second order perturbation theory, classical Poincar\'e surface
of section analysis of the intramanifold dynamics, and statistical analysis
of the intramanifold nearest-neighbor-spacing distributions.
For mutual orientations around $\beta\approx 60^\circ$ and excitation 
energies around the Stark saddle point the exact quantum calculation 
reveal large avoided crossings between states with the same approximate 
principal quantum number, $n$.
Because the level repulsion occurs preferably within a given $n$-manifold
it is interpreted as a quantum manifestation of {\em intramanifold chaos}.
This interpretation is strongly supported by the analysis of both the 
classical intramanifold dynamics and the intramanifold NNS distributions.
The classical analysis exhibits large stochastic regions, i.e., chaotic 
motion in the PSOS at those parameters where strong level repulsion has
been observed in the quantum spectra, and
the intramanifold NNS distributions show a transition from a Poisson
distribution to a Brody distribution at those energy-field parameters.

In parallel magnetic and electric fields ($\beta=0$) the intramanifold
dynamics is regular because of the cylindrical symmetry of the system and
the conservation of the $z$-component of the angular momentum.
It might be expected that intramanifold chaos increases the more the 
cylindrical symmetry is broken and becomes strongest in perpendicular 
crossed fields.
The interesting result of this Paper is that this assumption is not true
and intramanifold chaos becomes strongest, in our examples, at mutual
field orientations around $\beta\approx 60^\circ$.
Effects of intramanifold chaos are surprisingly weak at $\beta=90^\circ$
as can be seen in Figs.\ 8e and 10e.
Probably this cannot be explained by the existence of a discrete symmetry,
i.e., the $z$-parity, $\pi_z$, in perpendicular crossed fields and further 
investigations are necessary to understand the mechanism of the increase
and decrease of intramanifold chaos.

Implications are that some effects which have been discussed recently
for the crossed field atom such as Ericson fluctuations of the continuous 
photo cross section above threshold \cite{Mai92} and Arnol'd diffusion 
of the three-dimensional classical motion \cite{Mil96} might be
much stronger at mutual field orientations around $\beta\approx 60^\circ$
than in perpendicular crossed fields.
It might be useful to revisit these problems and extend the previous 
investigations to arbitrary mutual orientations of the external fields.

Up to now experimental investigation on atoms in combined electric and
magnetic fields have concentrated on parallel and perpendicular field
orientation.
If indeed some interesting physical effects are much stronger for special 
mutual field orientations than in parallel or perpendicular fields it
is a challenge to observe these effects experimentally.

\acknowledgements
We thank T.\ Uzer and J.\ von Milczewski for stimulating discussions.
This work was supported by the Deutsche Forschungsgemeinschaft (SFB 237).

\end{document}